\documentclass[prb,twocolumn,amsmath,amssymb,floatfix]{revtex4}
\usepackage{graphicx}
\usepackage{bm}

\newcommand{\be}{\begin{equation}}
\newcommand{\ee}{\end{equation}}
\newcommand{\beq}{\begin{eqnarray}}
\newcommand{\eeq}{\end{eqnarray}}

\begin{document}

\title{Universal adiabatic dynamics in the vicinity of a quantum critical point.}

\author{Anatoli Polkovnikov$^{1,2}$}
\affiliation{$^1$ Department of Physics, Boston University, 590
Commonwealth Ave., Boston, MA, 02215,\\ $^2$Department of Physics,
Harvard University, 17 Oxford St., Cambridge, MA 02138.}

\date{\today}

\begin{abstract}
We study temporal behavior of a quantum system under a slow external
perturbation, which drives the system across a second order quantum
phase transition. It is shown that despite the conventional
adiabaticity conditions are always violated near the critical point,
the number of created excitations still goes to zero in the limit of
infinitesimally slow variation of the tuning parameter. It scales
with the adiabaticity parameter as a power related to the critical
exponents $z$ and $\nu$ characterizing the phase transition. We
support general arguments by direct calculations for the Boson
Hubbard and the transverse field Ising models.
\end{abstract}

\maketitle

Quantum phase transitions have attracted a lot of theoretical and
experimental attention in recent decades, see for example
Ref.~[\onlinecite{Sachdev_book}]. They are driven entirely by
quantum fluctuations and occur at zero temperature. In this paper we
will be interested in second order transitions, which are
characterized by universal properties near the critical point.
Usually these properties can be revealed experimentally by measuring
various correlation functions. Since the relaxation time in most of
conventional condensed matter systems is relatively short, only
equilibrium or steady state regimes are experimentally relevant. On
the other hand recent progress in the realization of ultra cold
atomic gases~\cite{nature} made it possible to study experimentally
both equilibrium and strongly out of equilibrium properties of the
interacting quantum systems. Thus observation of the
superfluid-to-insulator transition~\cite{Bloch} relied on the
reversibility of the phase coherence after the system was slowly
driven to the insulating state and then back. In the same experiment
another resonant feature was observed if the Mott insulator is a
subject to an external linear potential of a particular strength.
This feature was later interpreted later as an Ising-like quantum
phase transition between normal and dipolar states~\cite{Sachdev2}.
The other big advantage of atomic systems is that the parameters
governing the transition can be tuned continuously during a single
experiment, so that, for example, it is possible to cross a quantum
critical point in a real time.

Let us consider now a specific situation, where some system was
initially in the ground state. Then a tuning parameter was slowly
changed to drive it through a critical point. From general
principles we know that the system should remain in the ground state
as long as it is protected by the gap from the excitations. On the
other hand the gap vanishes right at the critical point so the
adiabaticity conditions can never be satisfied in the vicinity of
the phase transition. The slower the parameter changes the more time
the system spends near the critical point, but on the other hand the
less the interval where the adiabaticity is violated. The
competition between these two processes determines the total amount
of excitations in the system. Here we show that the number of
excited states decreases as a power law of the tuning rate. Because
of the universality and scaling, below a certain dimension, which we
identify later, this power is determined by the critical exponents
$z$ and $\nu$ characterizing the transition. So measuring the number
of excitations as a function of the tuning rate one can obtain the
information about the critical properties of the phase transition.
We give a general argument for the particular scaling form and
consider two specific examples of phase transitions occurring within
Boson Hubbard- and transverse field Ising models, which confirm this
scaling.

Let us start from a general formalism. We assume that the system
is described by some Hamiltonian $\cal H(\lambda)$, which depends
on the external parameter $\lambda$. Without loss of generality
$\lambda=0$ corresponds to the phase boundary, so that $\lambda>0$
and $\lambda<0$ describe different phases of the system. Let the
set of (many-body) functions $\phi_r(\lambda)$ represent the
eigen-basis of the Hamiltonian $\cal H$. The wavefunction of the
system can be always expanded in this basis:
\be
\psi=\sum_p a_p(t) \phi_p(\lambda).
\label{eq 1}
\ee
We assume that $\lambda$ slowly changes in time:
$\lambda(t)=\delta t$, where $\delta$ is the adiabaticity
parameter and we took linear dependence on time for the sake of
convenience. Then substituting (\ref{eq 1}) into Schr\"odinger
equation we find:
\be
i {d a_p\over dt}+i\delta \sum_q a_q(t) \langle p|{d\over
d\lambda} |q\rangle = \omega_p(\lambda) a_p(t),
\label{eq 2}
\ee
where $\omega_p(\lambda)$ is the eigen frequency of the
Hamiltonian $\cal H(\lambda)$. It is convenient to perform a
unitary transformation:
\be
a_p(t)=\tilde a_p(t) \mathrm e^{-i \int^t \omega_p(\lambda(t))dt}=
\tilde a_p(\lambda)\mathrm e^{-{i\over \delta}\int^\lambda
\omega_p(\lambda)d\lambda}.
\label{eq 3}
\ee
Combining (\ref{eq 2}) and (\ref{eq 3}) we derive:
\be
{d \tilde a_p\over d\lambda}=-\sum_q \tilde a_q(\lambda)\langle
p|{d\over d\lambda} |q\rangle\, \mathrm e^{{i\over\delta}
\int^\lambda
(\omega_p(\lambda^\prime)-\omega_q(\lambda^\prime))d\lambda^\prime}.
\label{eq 4}
\ee
If before the evolution the system was in the ground state
$|0\rangle$ then a single term dominates the sum in (\ref{eq 4}).
The relative number of the excited states is thus given by:
\be
n_{ex}\approx\sum^\prime_p \left|\int_{-\infty}^\infty d\lambda
\langle p|{d\over d\lambda} |0\rangle\, \mathrm e^{{i\over\delta}
\int^\lambda
(\omega_p(\lambda^\prime)-\omega_0(\lambda^\prime))d\lambda^\prime}
\right|^2,
\label{eq 5}
\ee
where the prime over the sum implies that the summation is taken
only over the excited states. It is important to emphasize that
across the second order phase transition, which we consider in this
paper, the basis wave functions change continuously with $\lambda$.

Let us assume that we deal with a uniform d-dimensional system. This
assumption is not necessary, but it is the case for the most known
systems undergoing a quantum phase transition. We also assume that
there is a single branch of excitations characterized by the gap
$\Delta$ and some dispersion. Since both $\Delta$ and $\lambda$
become zero at the phase boundary then, near the critical point, we
must have $ \Delta\propto
|\lambda|^{z\nu}$~[\onlinecite{Sachdev_book}] with $z$ and $\nu$
being critical exponents. In the momentum space (\ref{eq 5}) reduces
to:
\be
n_{ex}\approx \int {d^d k\over
(2\pi)^d}\left|\int_{-\infty}^\infty d\lambda \langle k|{d\over
d\lambda} |0\rangle\, \mathrm e^{{i\over\delta} \int^\lambda
(\omega_k(\lambda^\prime)-\omega_0(\lambda^\prime))d\lambda^\prime}
\right|^2,
\label{eq 6}
\ee
From general scaling arguments we can write:
\be
\omega_k-\omega_0=\Delta F(\Delta/k^z)=\lambda^{z\nu} \tilde
F(\lambda^{z\nu}/k^z),
\label{eq 8}
\ee
where $F$ ($\tilde F$) is some undefined function satisfying
$F(x)\propto 1/x$ for large $x$, and $z$ is the dynamic critical
exponent. Similarly we can argue that:
\be
\langle k|{\partial\over \partial \Delta}|0\rangle = {1\over
k^z}G(\Delta/k^z)\Rightarrow \langle k|{\partial\over
\partial \lambda}|0\rangle = {\lambda^{z\nu-1}\over k^z}\tilde G(\lambda^{z\nu}/k^z),
\label{eq 9}
\ee
where $G$ ($\tilde G$) is another scaling function satisfying
$G(0)={\rm const}$. Having these scaling forms in mind we can do
the following substitutions in (\ref{eq 6}): $\lambda = k^{1/\nu}
\xi,\; k=\delta^{\nu\over z\nu+1}\eta.$ It is easy to see that if
the momentum integral in (\ref{eq 6}) can be extended to infinity
then:
\be
n_{ex}=C\delta^{d\nu\over z\nu+1},
\label{eq 11}
\ee
where $C$ is a nonuniversal constant which depends on the details of
the transition. The condition allowing to send the upper cutoff to
infinity is $d<d_c=2z(z\nu+1)$, where $d_c$ is the upper critical
dimension for this problem. Note that $\nu$ and $z$ can depend on
$d$ themselves. For $d>d_c$, the main contribution to the
excitations comes from the high momentum states. In this case
$n_{ex}$ would still vanish at $\delta\to 0$, but the universality
will disappear as excitations with high momenta ($k^z\gg \Delta$)
will dominate the transitions. The result (\ref{eq 11}) is quite
remarkable, it shows that if $\delta\to 0$ and $\nu$ is a finite
number greater then zero, the transitions to the excited states are
suppressed and the adiabatic limit still holds. We want to
emphasize, the adiabaticity is  understood in a sense that the
density, not the total number, of excitations is much smaller than
one. Strictly speaking in the true adiabatic limit there are should
be no excitations and the system must remain in the ground state.
However to achieve this, it is necessary to scale $\delta$ as
inverse power of the system size~\cite{xxx}, which is virtually
impossible to do in large systems. There are two limits $\nu\to 0$
and $\nu\to\infty$ in (\ref{eq 11}) which require special attention.
The first one is trivial since it corresponds to the absence of the
phase transition since the gap always remains finite except for a
very narrow interval around $\lambda=0$. Indeed, a more careful
analysis shows that the constant $C$ in (\ref{eq 11}) is
proportional to $\nu^2$. The opposite limit $\nu\to\infty$ is more
interesting since it corresponds to e.g. Kosterlits-Thouless (KT)
transition~\cite{Chaikin-Lubensky}, which has many realizations in
$1+1$ dimensional quantum systems. Thus if the precise scaling form
is $\Delta\propto \mathrm e^{-{b\over \lambda^r}}$ then
\be
n_{ex}\propto \delta^{d\over z}\ln^{{r+1\over r}{d-2z\over
z}}(\delta^{-1}).
\label{eq 11a}
\ee
This expression acquires extra logarithmic corrections as compared
to (\ref{eq 11}). In particular, for the KT transition $r=1/2$ and
$z=1$ so that (\ref{eq 11a}) reduces to
\be
n_{ex}^{KT}\propto \delta^d \ln^{3(d-2)}(\delta^{-1}).
\ee
There are no logarithmic corrections in two dimensions. However,
the only physically relevant case where the KT transition can
occur in a quantum system at zero temperature corresponds to
$d=1$.

Qualitatively one can interpret (\ref{eq 11}) in a simple way. The
transitions to the excited states occur when the adiabaticity
conditions break down, i.e. when ${d\ln\Delta\over dt}\geq \Delta$.
From this one immediately finds that the time interval when the
transitions take place scales as: $ t\sim \delta^{-{z\nu\over
z\nu+1}}.$ The typical gap at this time scale is
\be
\Delta\sim (\delta t)^{z\nu}\sim \delta^{z\nu\over z\nu+1},
\label{delta}
\ee
which amounts to the available phase space $\Omega\sim k^d\sim
\Delta^{d\over z}\sim\delta^{d\nu\over z\nu +1}.$ Now if we use
the anzats that $ d\Delta \langle k|{\partial/
\partial \Delta}|0\rangle$ is a scale independent quantity
(see (\ref{eq 9})), then we immediately come to the conclusion that
this phase space determines the number of excited states $n_{ex}$ so
that we come to (\ref{eq 11}). This simple derivation above, in
fact, does not rely on the spatial homogeneity of the system. The
only information we need is the density of states of excitations
$\rho(\epsilon)$ at the energy scale determined by (\ref{delta}). So
in a general case instead of (\ref{eq 11}) we get:
\be
n_{ex}\propto \delta^{z\nu\over z\nu+1}\rho(\delta^{z\nu\over
z\nu+1}).
\ee
Notice that (\ref{eq 11}) contains only two critical exponents $\nu$
and $z$. So measuring the dependence $n_{ex}(\delta)$ and knowing
one of the exponents, say $z$, one can  immediately deduce other.

Let us apply these ideas to the superfluid-to-insulator transition
in a system of interacting bosons in a $d$-dimensional lattice at
commensurate filling~\cite{Fisher, Sachdev_book}. To describe the
excitations near the critical point we adopt a mean-field
Hamiltonian derived by Altman and Auerbach in
Refs.~[\onlinecite{ehud1, ehud2}]:
\beq
&&\mathcal H=2dJN \sum_{\bf k} \bigg\{\left(2u \cos \theta-\cos
2\theta+\epsilon_{\bf k} \cos^2 \theta \right)b_{1,{\bf
k}}^\dagger b_{1,{\bf k}}\nonumber\\
&&~~~-{1-\epsilon_{\bf k}\over 2}\cos^2\theta(b_{1,{\bf
k}}^\dagger b_{1,-{\bf k}}^\dagger+b_{1,{\bf k}}b_{1,{\bf
-k}})\nonumber\\
&&~~~+(2u\cos^2{\theta\over 2}+\sin^2 \theta-\cos^2 {\theta\over
2} +\epsilon_{\bf
k}\cos^2{\theta\over 2})\,b_{2,{\bf k}}^\dagger b_{2,{\bf k}}\nonumber\\
&&~~~+ {1-\epsilon_{\bf k}\over 2}\cos^2{\theta\over 2}(b_{2,{\bf
k}}^\dagger b_{2,-{\bf k}}^\dagger+b_{2,{\bf k}}b_{2,{\bf
-k}})\bigg\}.
\label{eq 15}
\eeq
Here $J$ is the tunneling constant, $N$ is the mean number of bosons
per site, which we assume to be a large integer for the sake of
simplicity, $ \epsilon_{\bf k}={1\over 2d}\sum_\delta \mathrm
1-e^{i{\bf k}\delta}$,  $\theta$ is the mean field angle
characterizing the phase. In particular $\theta=0$ corresponds to
the Mott phase, while in the superfluid regime $\cos\theta
\approx u$. The dimensionless interaction $u\equiv U/(4JdN)$ is
defined so that the transition occurs at $u=1$.

In the Mott side of the transition, $u>1$, both branches are
degenerate and the Hamiltonian (\ref{eq 15}) can be readily
diagonalized via the Bogoliubov's transformation:
\be
\beta_{m,{\bf k}}=\cosh\phi_{m,{\bf k}}\,b_{m,{\bf k}}-\sinh\phi_{m,{\bf k}}\,b_{m,-{\bf
k}}^\dagger,
\label{eq 17}
\ee
where it becomes
\be
\mathcal H=\sum_{m,{\bf k}}\omega_{m,{\bf k}}\beta_{m,{\bf
k}}^\dagger \beta_{m,{\bf k}}.
\label{eq 18}
\ee
The eigenfrequencies $\omega_{m,{\bf k}}$ and the angle
$\phi_{m,{\bf k}}$ read:
\be
\omega_{1,2,{\bf k}}=4dJN u \sqrt{{u-1\over u}+{\epsilon_{\bf
k}\over u}}\approx 4dJN\sqrt{\lambda+\epsilon_{\bf k}},
\label{eq 20}
\ee
\be
\tanh 2\phi_{1,2,\,{\bf k}}=\pm {1-\epsilon_{\bf k}\over
2\lambda+1+\epsilon_{\bf k}},
\ee
We have chosen $\lambda=u-1$ to be the parameter governing the
phase transition. Given Hamiltonian (\ref{eq 18}) and
transformations (\ref{eq 17}) it is easy to write down the ground
state wavefunction:
\be
|0\rangle =\prod_{m,{\bf k}}\cosh \phi_{m,{\bf k}}\mathrm e^{\tanh
\phi_{m,{\bf k}}b_{m,{\bf k}}^\dagger b_{m,-{\bf k}}^\dagger
}|Vac\rangle,
\ee
where $|Vac\rangle$ is the state with no $b$ particles. It is a
simple exercise to check that $\langle p|\partial/\partial\lambda
|0\rangle$ is nonzero only when two particles with opposite
momenta are excited, i.e.
\be
|p\rangle\equiv |m,{\bf k},{\bf -k}\rangle=b_{m,{\bf k}}^\dagger
b_{m,-{\bf k}}^\dagger |0\rangle.
\ee
Then it can be verified that
\be
\langle m,{\bf k},-{\bf
k}|{\partial\over\partial\lambda}|0\rangle=\mp {1\over
2}{1-\epsilon_{\bf k}\over (1+\lambda)(\lambda+\epsilon_{\bf
k})}\approx \mp {1\over 2(\lambda+\epsilon_{\bf k})},
\label{eq 24}
\ee
where we used the approximation that both $\lambda$ and
$\epsilon_{\bf k}$ are small near the phase transition. Note that
(\ref{eq 24}) satisfies the general scaling (\ref{eq 9}) with the
exponent $\nu=1/2$, in the same way the dispersion relation
(\ref{eq 20}) agrees with (\ref{eq 8}).

A similar analysis can be performed on the superfluid side. The
Hamiltonian (\ref{eq 15}) gives now two branches corresponding to
the amplitude and the phase modes:
\be
\omega_{1,\bf k}\approx 4dJN\sqrt{-\lambda+\epsilon_{\bf k}},\;
\omega_{2,\bf k}\approx 4dJN\sqrt{\epsilon_{\bf k}},
\label{eq 29}
\ee
which are characterized by the following angles of the
transformation (\ref{eq 17}):
\be
\tanh 2\phi_{1,\bf k}\approx {1-\epsilon_{\bf k}+2\lambda\over
1+\epsilon_{\bf k}},\; \tanh 2\phi_{2,\bf k}\approx
{2-2\epsilon_{\bf k}+\lambda\over 2+2\epsilon_{\bf k}+\lambda}.
\ee
Note that the parameter $\lambda$ is negative on the superfluid
side. The matrix elements for these two modes are:
\be
\langle 1,{\bf k},{\bf -k}|{\partial\over
\partial\lambda}|0\rangle\approx -{1\over 2(\epsilon_{\bf
k}-\lambda)},\;\langle 2,{\bf k},{\bf -k}|{\partial\over
\partial\lambda}|0\rangle\approx {1\over 4}.
\label{eq 31}
\ee
The excitations of the phase modes are suppressed as compared to the
amplitude ones. This can be also expected on the physical grounds,
i.e. by changing the parameter $\lambda$ or equivalently $u$ we
change the mass or the gap of the amplitude mode thus exciting it,
however there is no such a coupling mechanism for the phase mode.
Clearly the total number of the excited phase oscillations is not
determined by the properties of the critical point and thus is not
universal. This number scales as $n_{ex}\propto \delta^{d}$, i.e.
vanishes much faster with $\delta$ than the number of excitations to
the gapped mode. Besides, a typical experiment would use the phase
contrast as a measure of superfluidity~\cite{Bloch, Orzel}, which is
not strongly affected by low momentum phase excitations. Keeping
this in mind we calculate explicitly only the number of particle
pairs lost to the mode $1$, which is gapped on both sides of the
transition. Experimentally this number can be detected by getting
first from the superfluid to the Mott insulator and then returning
back to the superfluid regime and measuring the loss of the phase
contrast, or by measuring the number of created particle-hole pairs
in the insulating state. Performing the integration in (\ref{eq 6})
we find that in one, two and three dimensions the number of
excitations is
\beq
&&n_{ex}^{1D}\approx 0.348 \left({\delta\over JN}\right)^{1\over
3},\; n_{ex}^{2D}\approx 0.059 \left({\delta\over
JN}\right)^{2\over
3},\nonumber\\
&& n_{ex}^{3D}\approx 0.010 \left({\delta\over JN}\right),
\label{mott}
\eeq
respectively. We note that since $\nu=1/2$ within this meanfield
treatment and $z=1$ the upper critical dimension is $2z(z\nu+1)=3$
so the scaling (\ref{eq 11}) is valid in one and two dimensions.
However, this model has an additional symmetry, giving the same
prefactors of the gap and wavefunction dependence on $\lambda$ in
both superfluid and insulating phases (compare (\ref{eq 20}) with
(\ref{eq 29}) and (\ref{eq 24}) with (\ref{eq 31})). This symmetry
shifts the upper critical dimension to $d=4$, so the results
remain universal in all three spatial dimensions. As anticipated,
expressions (\ref{mott}) are consistent with (\ref{eq 11}).

The correct description of the superfluid-to-insulator transition
gives exponents different from the mean-field values used above.
The commensurate transition lies in the universality class of the
XY model in the $d+1$ dimensions with $z=1$ and $\nu=0.5$ for
$d=3$, $\nu\approx 0.67$ for $d=2$ ~\cite{Zinn-Justin}, and
$\nu=\infty$ in $d=1$~\cite{Chaikin-Lubensky} (more precisely the
universality class in the latter case is of the KT transition with
$\Delta\propto\exp(-b/\sqrt{\lambda})$). So that for these special
points equation (\ref{eq 11}) reduces to:
\beq
&&n_{ex}^{1D}\propto {\delta\over \ln^3(\delta^{-1})},\quad
n_{ex}^{2D}\propto \delta^{0.80},\quad n_{ex}^{3D}\propto \delta.
\eeq
In a generic point of the superfluid-insulator transition
corresponding to the non-commensurate filling
$z\nu=1$~\cite{Fisher} and since $z\geq 1$ the upper critical
dimension $d_c$ is always larger than $3$. So (\ref{eq 11})
reduces to: $n_{ex}\propto \delta^{{d\over 2}\nu}.$ Thus measuring
$n_{ex}(\delta)$ it is possible to observe the exponent $\nu$.
Unfortunately except for $3D$, where $\nu=1/2$ and hence
$n_{ex}\propto \delta^{3/4}$ , the precise value of $\nu$ (and
$z$) is not fixed but rather depends on the point where the
transition occurs~\cite{Ferreira}.

Another example we consider here is the transverse field Ising
model~\cite{Sachdev_book}, which is described by the Hamiltonian
\be
\mathcal H_I=-\sum_j g\sigma_j^x+\sigma_j^z\sigma_{j+1}^z,
\label{is1}
\ee
where $\sigma_x$ and $\sigma_z$ are the Pauli matrices. The
dimensionless coupling constant $g$ drives the system through a
critical point, which occurs at
$g_c=1$~[\onlinecite{Sachdev_book}] and which is characterized by
the critical exponents $z=\nu=1$. Using the Jordan-Wigner
transformation one can show that~(\ref{is1}) maps to the model of
free spinless fermions with the Hamiltonian
\be
\mathcal H_I=-\sum_j c_j^\dagger c_{j+1}+c_{j+1}^\dagger c_j
+c_j^\dagger c_{j+1}^\dagger +c_{j+1}c_j-2gc_{j}^\dagger c_j,
\label{is2}
\ee
which in turn can be diagonalized by the Bogoliubov's
transformation:
\be
c_k=\cos(\theta_k/2)\gamma_k+i\sin(\theta_k/2)\gamma_{-k}^\dagger.
\ee
Here $c_k$ is the Fourier transform of $c_j$ and the angle
$\theta_k$ is given by~\cite{Sachdev_book}
\be
\tan\theta_k={\sin(k)\over \cos(k)-g}.
\ee
In the diagonal form the Hamiltonian (\ref{is2}) reads
\be
\mathcal H_I=\sum_k \varepsilon_k \gamma_k^\dagger\gamma_k,
\label{hi}
\ee
where $\varepsilon_k=2\sqrt{1+g^2-2g\cos k}$. The ground state
wavefunction, which is the vacuum of (\ref{hi}) reads
\be
|\Omega\rangle=\prod_k
(\cos(\theta_k/2)+i\sin(\theta_k/2)c_k^\dagger
c_{-k}^\dagger)|0\rangle
\ee
where $|0\rangle$ is the state with no $c$-fermions. The excited
states have the form of
$\gamma_{k1}^\dagger\gamma_{k2}^\dagger\dots\gamma_{kn}^\dagger|\Omega\rangle$.
The natural choice of the tuning parameter $\lambda$ is
$\lambda=g-1$ which is proportional to the energy gap $\Delta$. It
is straightforward to verify that
\beq
{\partial\over\partial \lambda}|\Omega\rangle&=& {i\over 2}\sum_k
{\partial\theta_k\over \partial g}
\gamma_k^\dagger\gamma_{-k}^\dagger|\Omega\rangle\nonumber\\
&=&{i\over 2} \sum_k {\sin k\over 1+g^2-2g\cos
k}\gamma_k^\dagger\gamma_{-k}^\dagger |\Omega\rangle,
\eeq
which again corresponds to two particle excitations. This
expression is consistent with (\ref{eq 9}) with $G(0)= {i\over
2}$. Now using (\ref{eq 6}) we immediately find:
\be
n_{ex}\approx 0.18 \sqrt{\delta},
\label{eq 34}
\ee
which agrees with the general formula (\ref{eq 11}) given that
$d=\nu=z=1$.

In conclusion, we showed that if the system, originally in the
ground state, is slowly driven through a quantum critical point,
the number of excited states per unit volume goes to zero as a
power law of the tuning rate. The exponent is universal and is
determined by the critical properties of the transition if the
dimension is smaller then $d_{cr}=2z(z\nu+1)$. We provided some
general arguments and performed explicit calculations for the
superfluid-to-insulator transition within the Boson Hubbard model
and for the quantum phase transition  in the transverse field
Ising model.

Recently, two other papers appeared, which addressed a similar issue
of the number of created defects for a specific case of a transverse
field Ising model~\cite{Zoller, Jacek}. In particular, the authors
got the same scaling as in Eq. (34) but with a slightly smaller
numerical prefactor. The discrepancy comes from a more accurate
treatment of transition probabilities within the Landau-Zeener
formalism~\cite{Jacek1}.

The author would like to acknowledge useful discussions with
E.~Altman, E.~Demler, M.~Lukin, S.~Sachdev and M.~Vojta. This work
was supported by US NSF grants DMR-0233773, DMR-0231631 and by the
Harvard Materials Research Laboratory via grant DMR-0213805.

\end{document}